\documentclass[floats,floatfix,amssymb,prd,twocolumn,superscriptaddress,nofootinbib]{revtex4-1}

\usepackage{amssymb,amsmath,verbatim,mathtools,enumitem,etoolbox,graphicx,microtype,afterpage,bm}

\usepackage[utf8]{inputenc}
\usepackage{amsmath,amssymb}
\usepackage{amsfonts}
\usepackage{bm}
\usepackage{courier}
\usepackage{graphics}
\usepackage{float}
\usepackage{amsmath}
\usepackage{amssymb}
\usepackage[mathscr]{euscript}
\usepackage{float}
\usepackage[caption=false]{subfig}
\usepackage{lipsum}
\usepackage{mathtools}
\usepackage{multirow}
\usepackage{graphicx}
\usepackage{mathtools}
\usepackage{multirow}
\usepackage{graphicx}
\usepackage[usenames,dvipsnames,table,xcdraw]{xcolor}
\usepackage[colorlinks, pdfborder={0 0 0}]{hyperref}
\usepackage[nameinlink,capitalise]{cleveref}
\usepackage{tensor}
\usepackage{verbatim}
\usepackage{wrapfig}
\usepackage{soul} %for line on text

\graphicspath{{fig}}
\begin{document}
%%%
\title{Black hole superradiant instability for massive spin-2 fields}

\author{\'Oscar~J.~C.~Dias}
	\email{ojcd1r13@soton.ac.uk }
	\affiliation{STAG research centre \& Mathematical Sciences, Univ. of Southampton, Highfield Campus, UK}

  \author{Giuseppe~Lingetti}
	\email{giuseppe.lingetti@uniroma1.it}
	\affiliation{Dipartimento di Fisica, ``Sapienza" Università di Roma \& Sezione INFN Roma1, Piazzale Aldo Moro 5, 00185, Roma, Italy}

 \author{Paolo~Pani}
	\email{paolo.pani@uniroma1.it}
	\affiliation{Dipartimento di Fisica, ``Sapienza" Università di Roma \& Sezione INFN Roma1, Piazzale Aldo Moro 5, 00185, Roma, Italy}

 \author{Jorge~E.~Santos}
	\email{jss55@cam.ac.uk }
	\affiliation{DAMTP, Centre for Mathematical Sciences, University of Cambridge, Wilberforce Road, Cambridge CB3 0WA, UK}

\begin{abstract}
Due to coherent superradiant amplification, massive bosonic fields can trigger an instability in spinning black holes, tapping their energy and angular momentum and forming macroscopic Bose-Einstein condensates around them. This phenomenon produces gaps in the mass-spin distribution of astrophysical black holes, a continuous gravitational-wave signal
emitted by the condensate, and several environmental effects relevant for gravitational-wave astronomy and radio images of black holes.
While the spectrum of superradiantly unstable mode is known in great detail for massive scalar (spin-0) and vector (spin-1) perturbations, so far only approximated results were derived for the case of massive tensor (spin-2) fields, due to the nonseparability of the field equations.
Here, solving a system of ten elliptic partial differential equations, we close this program and compute  the spectrum of the most unstable modes of a massive spin-2 field for generic black-hole spin and boson mass, beyond the hydrogenic approximation and including the unique dipole mode that dominates the instability in the spin-2 case.
We find that the instability timescale for this mode is orders of magnitude shorter than for any other superradiant mode, yielding much stronger constraints on massive spin-2 fields. 
These results pave the way for phenomenological studies aimed at constraining 
beyond Standard Model scenarios, ultralight dark matter candidates, and extensions to General Relativity using gravitational-wave and electromagnetic observations, and have implications for the phase diagram of vacuum solutions of higher-dimensional gravity.
\end{abstract}

\maketitle

%%%%%%%%%%%%%%%%%%%%%%%%%%%%%%%%%%%%%%%%%%%%%%%%%%%%%%%%%%%%%%%%%%%%%%%%%%%%%
\noindent{{\bf{\em Introduction.}}}
%%%%%%%%%%%%%%%%%%%%%%%%%%%%%%%%%%%%%%%%%%%%%%%%%%%%%%%%%%%%%%%%%%%%%%%%%%%%%
%
Ultralight bosons (such as the QCD axion, axion-like particles, dark photons, etc~\cite{Arvanitaki:2009fg,Essig:2013lka,Marsh:2015xka,Hui:2016ltb}) are predicted in several beyond Standard Model scenarios~\cite{Jaeckel:2010ni,Essig:2013lka,Hui:2016ltb,Irastorza:2018dyq}, including extra 
dimensions and string theories, and are compelling dark matter candidates. 
Searching for these fields in the laboratory is challenging due to their typically weak coupling to baryonic matter, but they might produce striking effects around astrophysical black holes~(BHs)~\cite{Arvanitaki:2009fg,Arvanitaki:2014wva,Brito:2015oca,Arvanitaki:2016qwi,Brito:2017wnc,Brito:2017zvb,Palomba:2019vxe,Isi:2018pzk,Brito:2020lup}. 
This possibility is allowed by the superradiant instability of spinning BHs against massive bosonic excitations~\cite{Press:1972zz,Detweiler:1980uk,Cardoso:2004nk,Shlapentokh-Rothman:2013ysa}, which occurs whenever the frequency $\omega_R$ of the perturbation satisfies the superradiant condition $0<\omega_R<m \Omega_{\rm H}$,
where $\Omega_{\rm H}$ is the horizon angular velocity and $m$ is the
azimuthal quantum number of the unstable mode (see~\cite{Brito:2015oca} for an overview). 
As an order-of-magnitude estimate, for a boson with mass $m_b= \mu\hbar$,
the superradiant instability is most effective when
its Compton wavelength is comparable to the BH gravitational radius, i.e. when the 
gravitational coupling $\alpha\equiv M\mu={\cal O}(0.1)$ (in the geometrized $G=c=1$ units henceforth adopted). This translates into the optimal condition for the instability $m_b\sim 10^{-11}(M_\odot/M)\,{\rm eV}$ but, as we shall see, in certain cases the range of relevant masses for which the instability is efficient can encompass several orders of magnitude.
In the superradiant regime the BH spins down, transferring energy and angular momentum to a mostly dipolar ($m=1$) boson condensate until $\omega_R\sim \Omega_{\rm H}$ and the instability quenches off. The condensate is then primarily dissipated through the emission of almost monochromatic quadrupolar gravitational waves~\cite{Arvanitaki:2014wva,Brito:2014wla}, with
frequency set by the boson mass. This process continues for $m>1$ modes on longer timescales~\cite{Ficarra:2018rfu,Brito:2015oca}.

While the qualitative aspects of this phenomenon are valid for any boson regardless of its spin, a crucial ingredient is the instability timescale of the dominant unstable mode, which strongly depends on $\alpha$ and type of massive boson~\cite{Brito:2015oca}.
Scalar (spin-0) fields are the best studied case~\cite{Damour:1976kh,Detweiler:1980uk,Zouros:1979iw,Dolan:2007mj,Arvanitaki:2014wva,Arvanitaki:2016qwi,
Brito:2017wnc,Brito:2017zvb}, but recent years have witnessed a significant progress also for vector (spin-1) fields~\cite{Pani:2012vp,Pani:2012bp,Witek:2012tr,Endlich:2016jgc,East:2017mrj,East:2017ovw,Baryakhtar:2017ngi, 
East:2018glu,Frolov:2018ezx,Dolan:2018dqv,Siemonsen:2019ebd}.
Massive tensor (spin-2) perturbations are much less understood. In this case the superradiant instability has been studied only perturbatively using a semi-analytical approach~\cite{Pani:2013pma} to first order in the spin~\cite{Brito:2013wya} and analytically in the so-called Newtonian regime where $\alpha\ll1$~\cite{Brito:2020lup}. However, the former
approximation is inaccurate for most astrophysical BHs given their sizable spin~\cite{Brenneman:2011wz,Middleton:2015osa,LIGOScientific:2021djp}, while the latter approximation is inaccurate in the most relevant regime for the instability, $\alpha={\cal O}(0.1)$, and also fails to capture the ``special'' dipole mode found numerically in 
Ref.~\cite{Brito:2013wya}, which was conjectured to be the dominant one.

Here we close an important gap in the BH superradiance program by computing the superradiant instability of a Kerr BH against massive spin-2 fields without approximations and for any mode. 
We extend the methods developed in Refs.~\cite{Dias:2009iu,Dias:2010maa,Dias:2010eu,Dias:2010gk,Dias:2011jg,Dias:2011tj,Dias:2013sdc,Cardoso:2013pza,Dias:2014eua,Dias:2015nua,Dias:2015wqa,Dias:2021yju,Dias:2022oqm} (see~\cite{Dias:2015nua} for an overview) and write the linearized field equations as a coupled system of ten elliptic partial differential equations~(PDEs) with associated
boundary conditions yielding an eigenvalue problem in the frequency domain.

In particular we will show that the instability timescale of the special dipole mode is dramatically shorter than for any other superradiantly unstable mode, reaching timescales comparable to the typical BH ringdown~\cite{Berti:2009kk,Dias:2015wqa,Dias:2021yju,Dias:2022oqm} (as short as $\tau\sim 2\times 10^{-4}(M/M_\odot)\,{\rm s}$ for a highly-spinning BH), and being effective in a much wider region of the parameter space.

%%%%%%%%%%%%%%%%%%%%%%%%%%%%%%%%%%%%%%%%%%%%%%%%%%%%%%%%%%%%%%%%%%%%%%%%%%%
\noindent{{\bf{\em Setup.}}}
%%%%%%%%%%%%%%%%%%%%%%%%%%%%%%%%%%%%%%%%%%%%%%%%%%%%%%%%%%%%%%%%%%%%%%%%%%%%%
At variance with scalar and vector fields, the coupling of a massive spin-2 field to gravity is highly 
nontrivial~\cite{Hinterbichler:2011tt,deRham:2010kj,Hassan:2011hr,Hassan:2011zd,deRham:2014zqa}. 
Here we will consider the field equations for a spin-2 perturbation $H_{ab}$ on a Ricci-flat background which is taken to be the Kerr solution~\cite{Kerr:1963ud}, although our computation does not depend on the details of the 
background and should be valid also for other stationary and axisymmetric solutions that might exist in bimetric 
theories~\cite{Brito:2013xaa,Babichev:2015xha}. The perturbed equations, describing the propagation of five physical degrees of freedom, are~\cite{Brito:2013wya,Mazuet:2018ysa}
\begin{subequations}
\begin{align} 
	&  \Box H_{ab} + 2R_{abcd}H^{cd} -\mu ^{2} H_{ab}=0	\label{eq:spin2_eom}\,, \\
	&\nabla^{a} H_{ab}=0\,, \qquad {H^{a}}_{a}=0\,, \label{eq:spin2_constraints}
\end{align}
\end{subequations}%
where $\Box=\nabla_a\nabla^a$ and $R_{abcd}$ are the D'Alembert operator and Riemann tensor of the background, respectively, and the first two terms describe the familiar Lichnerowicz operator $(\Delta_L H)_{ab}$. 
We shall discuss the embedding of these equations in various nonlinear theories later on.

In the standard Boyer-Lindquist coordinates ($t,r,\theta,\phi$) for the Kerr metric,  $\partial_t$ and $\partial_\phi$ are Killing vector fields, so one can decompose $H_{ab}$ along those directions: $H_{ab}(t,r,\theta,\phi)=e^{-i\omega t} e^{im\phi} \tilde{H}_{ab}(r,\theta)$. The resulting field equations for $\tilde{H}_{ab}$ are a coupled system of ten PDEs (of which five are constraints arising from Eq.~\eqref{eq:spin2_constraints}) which, after imposing boundary conditions, yield a  eigenvalue problem in the (complex) frequency $\omega=\omega_R+i\,\omega_I$, for a given mode $m$ (details in the Appendix).

The system of PDEs is then discretized using a pseudospectral collocation grid on Gauss-Chebyshev-Lobbato points in the compactified directions $y\in[0,1]$ and $x\in[-1,1]$, defined as $r=r_+/(1-y^2)$ and $\cos\theta=x\sqrt{2-x^2}$, where $r_+=M(1+\sqrt{1-\chi^2})$ is the BH event horizon and $\chi\equiv J/M^2\leq1$ is the dimensionless BH angular momentum.
The eigenfrequencies and associated eigenfunctions are found through a Newton-Raphson root-finding algorithm after imposing suitable boundary conditions for unstable modes~\cite{Dias:2015nua,Dias:2010maa}. The latter impose an exponential decay at asymptotic infinity ($y\to1$), regularity of the perturbations in ingoing Eddington-Finkelstein coordinates at the horizon ($y\to0$), and regularity on the north and south poles of the two-sphere, which also requires $m$ to be an integer. More details about the formulation of the problem and numerical method are presented in the Appendix.

\begin{figure}[t]
    \centering
    \includegraphics[width=0.495\textwidth]{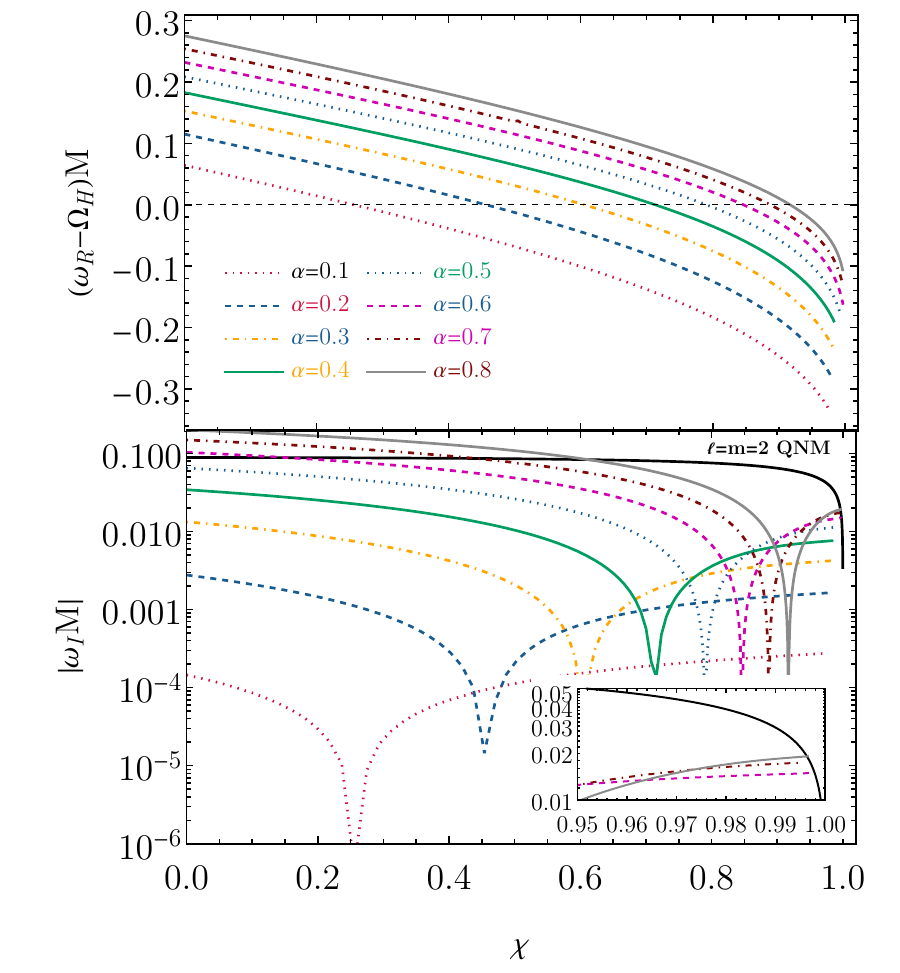}
    \caption{Real and imaginary parts of the dominant unstable dipole mode ($m=1$) of massive spin-2 perturbations of a Kerr BH as a function of the BH spin $\chi$. The unstable regime corresponds to $\omega_R<\Omega$, yielding $\omega_I>0$.
    For comparison, in the bottom panel we also show the (black) curve for the fundamental $\ell=m=2$ quasinormal mode of a Kerr BH~\cite{Berti:2009kk,Dias:2015wqa,Dias:2021yju,Dias:2022oqm}.
    The inset shows the comparison in the highly-spinning regime. In the Appendix we present three-dimensional and density plots obtained with 4018 grid points.
    }
    \label{fig:modes}
\end{figure}

%%%%%%%%%%%%%%%%%%%%%%%%%%%%%%%%%%%%%%%%%%%%%%%%%%%%%%%%%%%%%%%%%%%%%%%%%%%
\noindent{{\bf{\em Instability spectrum.}}}
%%%%%%%%%%%%%%%%%%%%%%%%%%%%%%%%%%%%%%%%%%%%%%%%%%%%%%%%%%%%%%%%%%%%%%%%%%%%%
Compared to the scalar and vector cases, massive spin-2 perturbations have two unique features, emerging already in the nonspinning case: i)~an unstable spherical ($m=0$) mode~\cite{Babichev:2013una,Brito:2013wya} which signals the existence of hairy BH solutions~\cite{Brito:2013xaa}; ii)~a special dipole polar mode~\cite{Brito:2013wya}, which does not fit within the general hydrogenic-like scaling found in the analytical Newtonian limit ($\alpha\ll1$). The latter predicts~\cite{Brito:2020lup}:
\begin{subequations}%
\begin{align}
  \frac{\omega_R}{\mu}&\simeq 1-\frac{\alpha^2}{2\left(\ell+n+S+1\right)^2}\,,\label{wR}\\
  \omega_I&\propto \alpha^{4\ell+5+2S}(\omega_R-m\Omega_{\rm H}), \label{wI}  
\end{align}
\end{subequations}%
where $\ell\geq0$, $n\geq0$, and the integer $S=(0,\pm1,\pm2)$ are the mode total angular momentum, overtone number, and polarization, respectively, with $m\in[-\ell,\ell]$. The sign change of $\omega_I$ due to the term $\omega_R-m\Omega_{\rm H}$ makes it clear that modes with $m\geq 1$ turn unstable in the superradiant regime, with an instability timescale $\tau=1/\omega_I$.
As predicted by~\eqref{wI}, the dominant hydrogenic spin-2 mode (with $\ell=2=-S$) has a 
% timescale comparable 
parametrically longer timescale compared 
to the dominant hydrogenic spin-1 mode (with $\ell=1=-S$)~\cite{Brito:2020lup}.
On the other hand, the special dipole ($m=1$) spin-2 mode computed numerically in the nonspinning case displays a different behavior~\cite{Brito:2013wya}:
%%%
%
\begin{equation}
\frac{\omega_R^{\rm dipole}}{\mu}\approx 0.72(1-\alpha)\,, \qquad \omega_I^{\rm dipole}\propto \alpha^{3}\mu\,. \label{wRwIdipole}
\end{equation}
Thus, in the nonspinning case this mode has the largest binding energy, $\omega_R/\mu-1$, and the shortest decay time of the entire spectrum for massive bosonic perturbations.

Our numerical method is general and applies to all unstable modes. Indeed, we have confirmed numerically the analytical results for the hydrogenic modes, Eqs.~\eqref{wR}--\eqref{wI}.
Henceforth we focus on the special dipole mode, since \cite{Brito:2013wya} found indication that $\omega_I^{\rm dipole}\propto \alpha^{3}(\omega_R-m\Omega_{\rm H})$ to leading order in the BH spin ($m=1$), suggesting the shortest instability timescale in the superradiant regime. 

In Fig.~\ref{fig:modes} we show the behavior of this special mode for generic BH spin $\chi$ and for a large range of gravitational coupling up to $\alpha=0.8$. While we confirm the results of Ref.~\cite{Brito:2013wya} in the nonspinning case, the most striking feature of this mode emerges in the regime that was not possible to explore so far, namely large BH spins and large coupling.
For nearly extremal BHs, we find that this mode has an imaginary part as large as $\omega_I M\approx 0.019$ when $\alpha\approx 0.8$. This translates into the instability timescale
\begin{equation}
    \tau \approx 2.6\times 10^{-4} \left(\frac{M}{M_\odot}\right)\,{\rm s}\,, \label{timescale}
\end{equation}
%%%
almost two orders of magnitude shorter than for any other superradiant mode known so far, including the dominant unstable mode in the massive spin-1 case~\cite{Dolan:2018dqv} and any other massive spin-2 modes.
%
% for spin-1: alpha~0.5 wI~4e-4
%
Interestingly, as shown by the inset in Fig.~\ref{fig:modes}, the dominant quasinormal mode of a highly-spinning Kerr BH~\cite{Berti:2009kk,Dias:2015wqa,Dias:2021yju,Dias:2022oqm} has a decay time comparable to, or even longer than, \eqref{timescale}.
This means that the instability is so fast that it would affect the ringdown of a newly formed BH.

Overall, our numerical data are described by a simple polynomial fit:  
%\begin{subequations}%
\begin{align}
\frac{\omega_R^{\rm dipole}}{\mu}&\approx \left(\sum_{i=0}^3 a_i \chi^i\right) \left(1+ \alpha\sum_{i=0}^3 b_i \chi^i +\alpha^2\sum_{i=0}^2 c_i \chi^i\right) \,,%\label{wRdipolefit} 
\nonumber \\
%%%
\omega_I^{\rm dipole}&\approx -\alpha^{3}(\omega_R-\Omega_{\rm H}) \sum_{i=0}^2 d_i \chi^i
\,,\label{wIdipolefit}
\end{align}
%\end{subequations}%
with $a_i\approx (0.73,-0.05,0.15,-0.12)$, $b_i=(-1.21,0.68,-0.55,0.61)$, $c_i=(0.69,-0.58,-0.11)$, $d_i=(1.47,1.86,-2.75)$.
In the unstable regime the fits of $\omega_R^{\rm dipole}$ and $\omega_I^{\rm dipole}$ are accurate within $2\%$ and $80\%$, respectively, in the range $\alpha\in[0.05,0.8]$ and $\chi\in[0,\approx0.99]$. 

\begin{figure*}[th]
\includegraphics[width=0.495\textwidth]{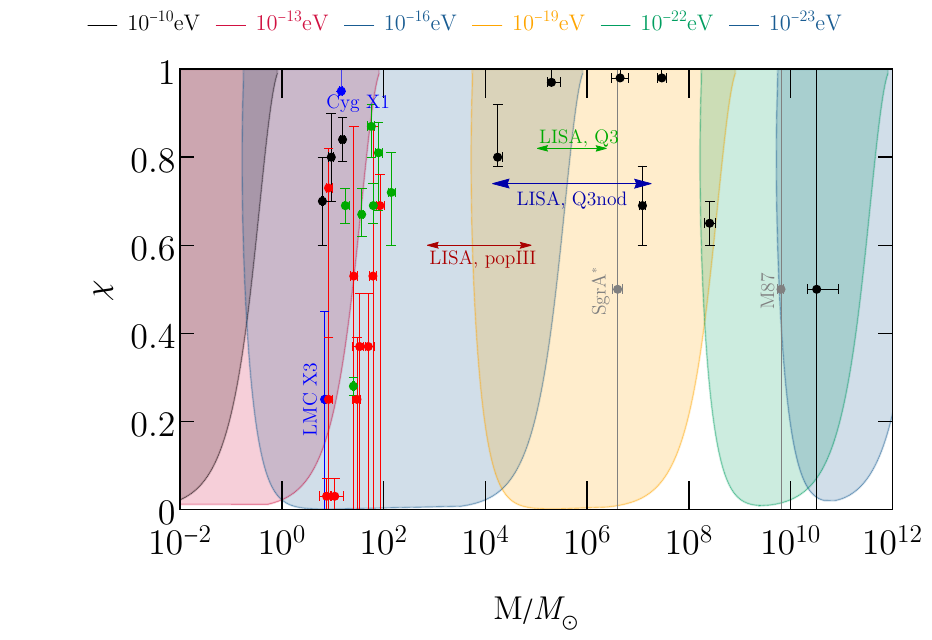}
\includegraphics[width=0.495\textwidth]{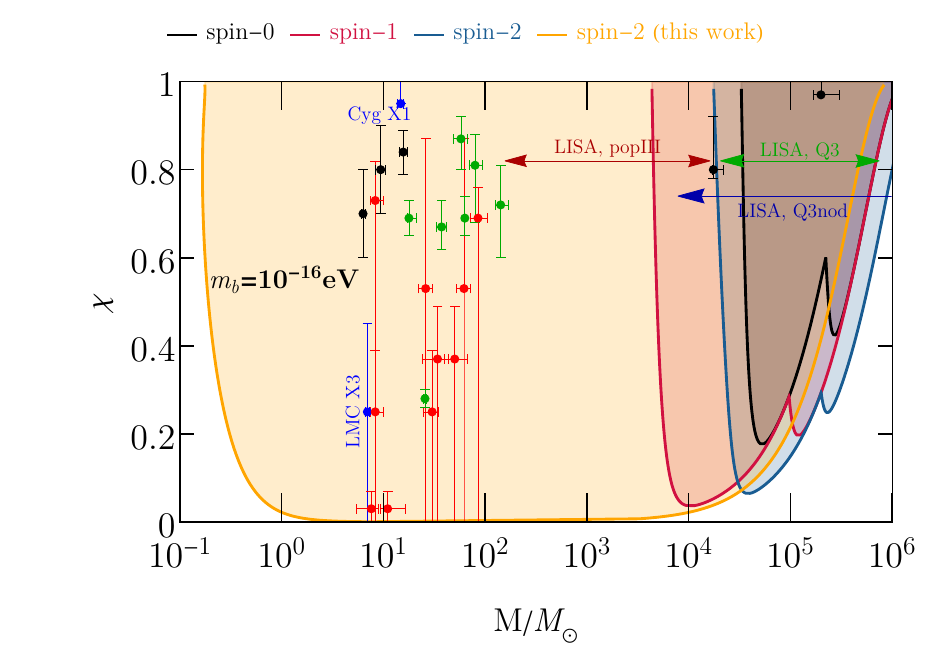}
\caption{\emph{Left:} Exclusion regions in the BH spin-mass diagram obtained from the superradiant 
instability of Kerr BHs against massive spin-2 fields for the dominant $m=1$ mode.
For each mass of the field, the separatrix corresponds to an instability timescale equal to the Salpeter time, $\tau_S=4.5\times 10^7 {\rm\, yr\,}$.
Markers and error bars are explained in the main text.
\emph{Right:}~Comparison of the exclusion region for a boson with mass $m_b=10^{-16}\,{\rm eV}$ for the dominant mode of scalar, vector, tensor perturbations. In the latter case we show the result computed in Ref.~\cite{Brito:2020lup} and the new mode computed in this work.
\label{fig:Regge}}
% \end{center}
\end{figure*}
%

%%%%%%%%%%%%%%%%%%%%%%%%%%%%%%%%%%%%%%%%%%%%%%%%%%%%%%%%%%%%%%%%%%%%%%%%%%%%%
\noindent{{\bf{\em BH mass-spin gaps.}}}
%%%%%%%%%%%%%%%%%%%%%%%%%%%%%%%%%%%%%%%%%%%%%%%%%%%%%%%
%
A generic prediction of superradiant instabilities is that highly-spinning BHs should lose angular momentum over a timescale $\tau=1/\omega_I$ that might be much shorter than typical astrophysical timescales.
Thus, if ultralight bosons exist in the Universe they would lead to statistical evidence for slowly rotating BHs in a region of the 
``Regge'' plane (mass versus angular momentum) of astrophysical 
BHs~\cite{Arvanitaki:2010sy,Brito:2015oca,Baryakhtar:2017ngi,Brito:2017wnc,Brito:2017zvb,Ng:2019jsx,Fernandez2019,Stott:2020gjj,Unal:2020jiy,Unal:2023yxt}. 
Given a $m_b$, there exists a forbidden region (a gap~\cite{Arvanitaki:2010sy}) in the Regge plane wherein the instability timescale is shorter than known spin-up astrophysical processes such as accretion. 

The exclusion regions in the Regge plane are shown in the left panel of Fig.~\ref{fig:Regge} for selected values of $m_b$, whereas in the right panel we compare the exclusion region obtained in this work with those of other superradiant modes for the same representative value $m_b=10^{-16}\,{\rm eV}$. Owing to the much shorter instability timescale, our exclusion region encompasses several orders of magnitude, ranging from stellar to supermassive BHs, for the same $m_b$. 

Data points shown in Fig.~\ref{fig:Regge} refer to several observations: 
{\bf (i)}~Black points denote electromagnetic estimates of the mass and spin of accreting stellar and supermassive BHs~\cite{Brenneman:2011wz,Middleton:2015osa}, conservatively requiring that $\tau<\tau_S=4.5\times 10^7 {\rm\, yr\,}$, where $\tau_S$ is the Salpeter timescale for accretion. We also include some recent candidates of spinning intermediate-mass BHs, although mass-spin measurements for these sources might be sensitive to modelling systematics~\cite{Wen:2021yhz,Cao:2022oic}.
{\bf (ii)}~Red points are the $90\%$ confidence levels for the
binary BHs in a selection of the merger events detected by the LIGO-Virgo-KAGRA collaboration so far~\cite{LIGOScientific:2018mvr,Venumadhav:2019lyq} (using the errors on the binary effective spin~\cite{PhysRevLett.106.241101} as a proxy for the errors on the individual spins).
While individual spin measurements coming from current gravitational-wave events have large errors, some binaries have confidently nonvanishing spins~\cite{LIGOScientific:2020kqk,LIGOScientific:2021djp} and future LIGO detections will provide measurements of the individual spins with $\approx30\%$ accuracy~\cite{TheLIGOScientific:2016pea}. 
Much more accurate spin measurements of binary BHs out to cosmological distances will come from the future LISA 
mission~\cite{Audley:2017drz} for intermediate and supermassive objects, mainly depending on the mass of BH seeds in the early Universe (the horizontal arrows in Fig.~\ref{fig:Regge} denote the range of projected 
mass measurements with LISA using three different population models for supermassive BHs~\cite{Klein:2015hvg,Brito:2017zvb}).
This would allow searching for ultralight bosons in a large mass range $m_b\in(10^{-19},10^{-13})\,{\rm eV}$, much wider than in previous cases~\cite{Brito:2017wnc,Brito:2017zvb,Cardoso:2018tly,Unal:2023yxt}.
{\bf (iii)}~Green points are the $90\%$ confidence levels for the mass-spin of a selection of the merger 
remnants~\cite{LIGOScientific:2018mvr,LIGOScientific:2020kqk,LIGOScientific:2021djp}, which also identify targets of merger follow-up 
searches~\cite{Arvanitaki:2014wva,Arvanitaki:2016qwi,Baryakhtar:2017ngi,Isi:2018pzk,Ghosh:2018gaw}.
In the particular case at hand, it is tantalizing that $|\omega_I|$ can be comparable to or even larger than the typical $|\omega_I|$ of a BH quasinormal mode~\cite{Berti:2009kk,Dias:2015wqa,Dias:2021yju,Dias:2022oqm} (see Fig.~\ref{fig:modes}).
This implies that the instability would directly affect the postmerger phase by dynamically reducing the spin of the remnant during the ringdown, which does not happen for any other superradiant mode. 
{\bf (iv)} 
The blue points refer to Cyg~X-1~\cite{Gou:2009ks} and LMC X-3~\cite{Steiner:2010kd}, for which there are reliable spin measurements suggesting that the spin is constant over ${\cal O}(10\,{\rm yr})$~\cite{Middleton:2015osa}. In these cases, more direct constraints come by imposing $\tau<{\cal O}(10\,{\rm yr})$~\cite{Cardoso:2018tly}.
Given the strong instability, these sources could confidently exclude spin-2 particles with mass $m_b\in(10^{-15},5\times 10^{-12})$.
{\bf (v)}~Finally, the two gray points correspond to the masses of SgrA$^*$ and M87 as measured by the Event Horizon 
Telescope~\cite{Akiyama:2019cqa,Akiyama:2019eap,EventHorizonTelescope:2022wkp}. While an accurate spin measurement for these sources is still not available, they are expected to have moderate to large 
spin~\cite{Akiyama:2019fyp,Tamburini:2019vrf,EventHorizonTelescope:2022wkp}. Given the extent of the exclusion regions in Fig.~\ref{fig:Regge}, any spin measurement in SgrA$^*$ and M87 that confidently excludes zero would approximately rule out the range $m_b\in[10^{-23},10^{-17}]\,{\rm eV}$, which is much wider than the exclusion regions derived with other modes~\cite{Davoudiasl:2019nlo,Chen:2019fsq,Cunha:2019ikd}.
Likewise, if some of the ultramassive BHs with $M\simeq {\rm few}
\times 10^{10} 
M_\odot$ \cite{McConnell:2011mu,2012arXiv1203.1620M,Nightingale:2023ini} (rightmost black data point in the left panel of Fig.~\ref{fig:Regge}) were confirmed to have nonzero spin~\cite{Riechers:2008xt}, it would be possible to exclude an ultralight spin-2 field with mass even slightly smaller than $m_b\approx 10^{-23}\,{\rm eV}$.

%%%%%%%%%%%%%%%%%%%%%%%%%%%%%%%%%%%%%%%%%%%%%%%%%%%%%%%%%%%%%%%%%%%%%%%%%%%%%
\noindent{{\bf{\em Embeddings of the model.}}} 
%%%%%%%%%%%%%%%%%%%%%%%%%%%%%%%%%%%%%%%%%%%%%%%%%%%%%%%%%%%%%%%%%%%%%%%%%%%%%
At the nonlinear level, there is a unique way to couple a dynamical spin-2 field to gravity~\cite{deRham:2010kj,Hassan:2011hr,Hassan:2011zd,Babichev:2016bxi}.
Nonetheless, Eqs.~\eqref{eq:spin2_eom}--\eqref{eq:spin2_constraints} emerge from linear perturbations in several different contexts.
In ghost-free massive gravity~\cite{deRham:2010kj}, these equations describe the linear dynamics of a massive graviton when considering a Ricci-flat reference metric.
In the context of bimetric theories~\cite{Hassan:2011hr,Hassan:2011zd}, they describe the dynamics of perturbations propagating on a fixed background given by two copies of the same Ricci-flat metric. 

Interestingly, Eqs.~\eqref{eq:spin2_eom}--\eqref{eq:spin2_constraints} also arise from BH perturbations in Einstein-Weyl gravity~\cite{PhysRevD.32.379} 
%%%
\begin{equation}
    S=\int d^4x\sqrt{-g}\left(R-\frac{\beta}{6} C_{abcd}C^{abcd}\right)\,,\label{EinsteinWeyl}
\end{equation}
%%%
where $C_{abcd}$ is the Weyl tensor. This theory propagates a massless graviton and a massive spin-2 field and (although the latter has ghosts~\cite{PhysRevD.16.953}) it has a number of unique features: it is renormalizable in Minkowski spacetime~\cite{PhysRevD.16.953}, admits all solutions of vacuum General Relativity, and has a well-posed initial value problem even if the corresponding field equations are of fourth order~\cite{Noakes:1983xd}.
By linearizing the theory on a Ricci-flat background~\cite{Myung:2013doa}, the equations for the perturbed Ricci tensor take the form~\eqref{eq:spin2_eom}--\eqref{eq:spin2_constraints}
upon identification $H_{ab}\equiv \delta R_{ab}$ and $\mu^2 \equiv 1/\beta$.
%\begin{equation} \mu^2 = \frac{1}{\beta}\,. \end{equation}
Thus, in the strongly-coupled regime ($\beta\to\infty$) the theory propagates only a massive graviton with small effective mass, and our bounds on $\mu$ can be directly translated into bounds on the coupling $\beta$, which are notoriously very hard to place.

Finally, Eqs.~\eqref{eq:spin2_eom}--\eqref{eq:spin2_constraints} emerge also in the context of higher-dimensional gravity when studying linear perturbations 
of a five-dimensional ($d=5$) black string --~obtained by adding a flat direction to the Schwarzschild or Kerr BH~-- and performing a Fourier decomposition with Fourier momentum $k\equiv \mu$ along the extra dimension~\cite{Gregory:1993vy}.
For example, the aforementioned spherical ($m=0$) unstable mode~\cite{Babichev:2013una}, present for $\alpha\leq \alpha_{\rm crit}\approx 0.438$ in the static case, corresponds to the familiar Gregory-Laflamme instability of a black string under fragmentation~\cite{Gregory:1993vy,Lehner:2010pn,Figueras:2022zkg}. The onset of this instability in $d=5$ signals a new branch of nonuniform black strings~\cite{Gubser:2001ac,Harmark:2002tr,Kol:2002xz,Wiseman:2002zc,Kol:2003ja,Harmark:2003dg,Harmark:2003yz,Kudoh:2003ki,Sorkin:2004qq,Gorbonos:2004uc,Kudoh:2004hs,Dias:2007hg,Harmark:2007md,Wiseman:2011by,Figueras:2012xj,Dias:2017coo,Dias:2022mde,Dias:2022str,Dias:2023nbj} which corresponds, in $d=4$, to hairy BHs of massive gravity~\cite{Brito:2013xaa}.

A natural extension of our work is to track the $m=0$ mode in the spinning case to obtain accurate estimates for $\alpha_{\rm crit}(\chi)$~\cite{Monteiro:2009ke} and associated timescales. Given the large domain of existence and the short timescale of the special dipole mode, it would be relevant to assess whether there exists a range of $\alpha$ in which the $m=0$ instability is absent or weak, whereas the $m=1$ instability discussed here is dominant. Work in $d=6$ dimensions \cite{Dias:2022mde,Dias:2022str,Dias:2023nbj} suggests that such a window might indeed exist.

In all nonlinear completions aforementioned, the special dipole mode is marginally stable ($\omega_I=0$) when the superradiant condition is saturated, $\omega=\Omega_H$ (as a further check of our code we confirmed this feature to very high accuracy).
A purely real mode at the linear level strongly suggests the existence of a nonlinear stationary solution branching off the critical point~\cite{Dias:2011at,Herdeiro:2014goa,Dias:2015nua,Ishii:2018oms,Ishii:2019wfs}.
For special black strings in $d=6$ (with enhanced symmetries that reduce Eqs.~\eqref{eq:spin2_eom}--\eqref{eq:spin2_constraints} to ODEs) the superradiant onset signals the existence of new rotating black strings -- denoted as resonator~\cite{Dias:2022str} and helical~\cite{Dias:2023nbj}  black strings -- that are time-periodic (but not time independent neither axisymmetric) since they have a helical Killing vector field.
%These solutions were denoted as helical black strings~\cite{Dias:2023nbj} (they preserve translations along $z$) and black resonator strings~\cite{Dias:2022str} (they break $z$-translations). 
Returning to $d=5$, the instability onset in Fig.~\ref{fig:modes} is similarly expected to signal black resonator/helical strings bifurcating from the Kerr string in a phase diagram of solutions.
From the $d=4$ perspective, this should correspond to the existence of novel hairy rotating BHs of massive/bimetric/Einstein-Weyl gravity with a single helical Killing vector field that would have higher entropy than the Kerr BH.
It would be interesting to find such solutions which are expected to have properties very different from the hairy BHs with scalar/vector hair~\cite{Herdeiro:2014goa,Herdeiro:2016tmi} (see~\cite{Herdeiro:2015waa} for a review), as it occurs for systems in anti-de Sitter with similar superradiant physics~\cite{Dias:2011at,Cardoso:2013pza,Horowitz:2014hja,Dias:2015rxy}. 
Yet, when emerging dynamically as the end-state or metastable state of the instability, these solutions would have an angular momentum close to the superradiant threshold, so our bounds derived from BH mass-spin measurements are robust.

%%%%%%%%%%%%%%%%%%%%%%%%%%%%%%%%%%%%%%%%%%%%%%%%%%%%%%%%%%%%%%%%%%%%%%%%%%%%%
\noindent{{\bf{\em Conclusion.}}}
%%%%%%%%%%%%%%%%%%%%%%%%%%%%%%%%%%%%%%%%%%%%%%%%%%%%%%%%%%%%%%%%%%%%%%%%%%%%%
We have shown that the special dipole spin-2 mode of a spinning BH does not fit in the standard hydrogenic-like picture of superradiantly unstable modes of massive bosonic perturbations. Most importantly, it has by far the shortest instability timescale among all superradiant modes.
This has several direct consequences which are unique to massive spin-2 fields. First of all, in the highly-spinning case the instability timescale can be comparable to or even larger than the dominant quasinormal mode of a Kerr BH. This indicates that the instability can be effective during the ringdown in a post-merger phase, and suggests the need for novel ringdown-based searches accounting for this effect.
Furthermore, the exclusion regions in the BH mass-spin diagram are much wider than for other superradiant modes. In particular, the same ultralight boson mass (around $m_b\approx 10^{-16}\,{\rm eV}$) would give rise to interesting phenomenology for both stellar-mass and supermassive BHs, facilitating possible multiband constraints on the model.
The exclusion regions for each $m_b$ extend almost up to $\chi\approx0$, showing that \emph{any} BH spin measurement can be translated into interesting constraints. 
Owing to the wideness of the BH mass-spin gaps, the range of detectable spin-$2$ masses is much larger than in all other cases explored so far.
For example, current observations of stellar-mass BHs (either in the electromagnetic or in the gravitational-wave band) exclude the region $m_b\in [ 5\times10^{-17},5\times10^{-11}]\,{\rm eV}$, which already severely constrains the observability of dipolar radiation from BH binaries in these models~\cite{Cardoso:2018zhm,dipoleinprep}.
Overall, few BH mass-spin measurements distributed in the region $M\sim M_\odot$ to $M\sim 10^{10} M_\odot$ would probe massive spin-2 fields in the impressive range $m_b\in[ 10^{-23},10^{-10}]\,{\rm eV}$, which also includes $m_b\sim 10^{-22}\,{\rm eV}$, where ultralight bosons are compelling dark-matter candidates~\cite{Hui:2016ltb}. To the best of our knowledge, there are no other constraints on ultralight spin-2 fields in this mass range, so superradiant instabilities provide us with a unique discovery opportunity.

A natural extension of our work is to use these new unstable modes to compute the rich phenomenology associated with the gravitational-wave emission from the bosonic condensate~\cite{Arvanitaki:2016qwi,Baryakhtar:2017ngi,Brito:2017wnc,Brito:2017zvb,Brito:2020lup,Isi:2018pzk,Ghosh:2018gaw,Tsukada:2018mbp,Palomba:2019vxe,Sun:2019mqb}, as well as  environmental effects of the condensate in BH binary inspirals~\cite{Hannuksela:2018izj,Baumann:2018vus,Zhang:2018kib,Berti:2019wnn,Baumann:2019eav,Baumann:2019ztm,Zhang:2019eid,Cardoso:2020hca,DeLuca:2021ite,Baumann:2022pkl} and for radio images~\cite{Davoudiasl:2019nlo,Chen:2019fsq,Cunha:2019ikd}.
Due to the stronger instability, these effects are anticipated to be much more prominent than in other cases studied so far.

%%%%%%%%%%%%%%%%%%%%%%%%%%%%%%%%%%%%%%%%%%%%%%%%%%%%%%%%%%%%%%%%%%%%%%%%%%%
\noindent{{\bf{\em Acknowledgments.}}}
%%%%%%%%%%%%%%%%%%%%%%%%%%%%%%%%%%%%%%%%%%%%%%%%%%%%%%%%%%%%%%%%%%%%%%%%%%%%%
 We thank Vitor Cardoso,  Francisco Duque, Andrea Maselli, and David Pereñiguez for interesting comments on the draft.
O.~D. acknowledges financial support from the STFC ``Particle Physics Grants Panel (PPGP) 2018" Grant No.~ST/T000775/1. O.D.'s research was also supported in part by the International Centre for Theoretical Sciences (ICTS), India, in association with the program "Nonperturbative and Numerical Approaches to Quantum Gravity, String Theory and Holography" (code: ICTS/numstrings-2022/8).
G.L. and P.P. acknowledge financial support provided under the European
Union's H2020 ERC, Starting Grant agreement no.~DarkGRA--757480, under
MIUR PRIN (Grant 2020KR4KN2 “String Theory as a bridge between Gauge Theories and Quantum Gravity”) and FARE (GW-NEXT, CUP: B84I20000100001, 2020KR4KN2) programmes, and support from the Amaldi Research Center funded by the MIUR program ``Dipartimento di Eccellenza" (CUP:~B81I18001170001). This work was supported by the EU Horizon 2020 Research and Innovation Programme under the Marie Sklodowska-Curie Grant Agreement No. 101007855.  J.E.S. has been partially supported by STFC consolidated grant ST/T000694/1. The authors acknowledge the use of the IRIDIS High Performance Computing Facility, and associated support services at the University of Southampton, for the completion of this work.

\appendix

\section{Details on the eigenvalue problem and numerical methods}\label{app:1}

Here we describe in detail how we solve the perturbed equations, Eqs.~(1a)-(1b) of the main text, around the Kerr BH background whose gravitational field is
\begin{align}\label{metric}
ds^2=&-\frac{\Delta}{\Sigma}\left[\mathrm{d}t-a\sin^2\theta\,\mathrm{d}\phi \right]^2
+\frac{\Sigma}{\Delta}\,\mathrm{d}r^2+\Sigma\,\mathrm{d}\theta^2 \nonumber\\
& +\frac{\sin^2\theta}{\Sigma}\left[(r^2+a^2)\mathrm{d}\phi -a\,\mathrm{d}t\right]^2 \,,
\end{align}
where
\begin{equation} \label{metricAux}
\Delta=r^2+a^2-2M r\,,\qquad\qquad \Sigma=r^2+a^2 \cos^2\theta\,.
\end{equation}
The event horizon is a null hypersurface with $r=r_+$, with $r_+$ being the largest positive real root of $\Delta$. Furthermore, $M=\frac{r_+^2+a^2}{2r_+}$ is the ADM mass of the BH, $J=M a$ is its ADM angular momentum, and $|a|\leq M$, with equality saturating at extremality, where the BH event horizon becomes degenerate. The temperature and angular velocity of the Kerr solution are 
\begin{equation} \label{KerrTOm}
T_H=\frac{r_+^2-a^2}{4\pi r_+(r_+^2+a^2)}\,,\qquad\qquad \Omega_H=\frac{a}{r^2+a^2}\,.
\end{equation}

Since $\partial_t$ and $\partial_\phi$ are Killing vector fields of \eqref{metric},  we can decompose the perturbations $H_{ab}$ along those directions: $H_{ab}(t,r,\theta,\phi)=e^{-i\omega t} e^{im\phi} \tilde{H}_{ab}(r,\theta)$ where we are interested on the $m=1$ case (onwards our discussion applies to this case). 
Moreover, we find it convenient to work with a compact radial coordinate $y\in[0,1]$ and with a new polar coordinate  $x\in[-1,1]$ related to the standard Boyer-Lindquist coordinates $(r,\theta)$ by
\begin{equation} \label{coordtransf}
r=\frac{r_+}{1-y^2}\,,\qquad \cos\theta=x\sqrt{2-x^2}\,.
\end{equation}
It is further useful to work with the dimensionless quantities\footnote{Our final results are presented in the main text in terms of the more natural dimensionless quantities $\chi\equiv J/M^2=a/M$, $\omega \,M$ and $\alpha \equiv \mu\, M$.}  
\begin{align} \label{dimensionless}
&\widetilde{a}=\frac{a}{r_+}\,,\quad \widetilde{\omega}=\omega\, r_+\,,\quad \widetilde{\mu}=\mu\, r_+\,, \nonumber \\
&\widetilde{T}_H=T_H r_+\,,\quad \widetilde{\Omega}_H=\Omega_H r_+\,.
\end{align}
In these conditions, the perturbed equations (1a)-(1b) in the main text  describe a coupled system of ten PDEs for $\tilde{H}_{ab}(y,x)$, where five of them are the  transverse and the  traceless constraints~(1b), as detailed next. The traceless condition gives an algebraic equation for $\tilde{H}_{tt}$ as a function of $\{ \tilde{H}_{yy},\tilde{H}_{xx},\tilde{H}_{\phi\phi} \}$. This algebraic equation together with other nine equations constitute a minimal set of equations that close the full system of perturbed equations. These nine equations are the four transverse conditions in~(1b) and the five  components $\{xx,xy,x\phi,yy,y\phi \}$ of (1a).
The ten unknown functions are $\tilde{H}_{ab}(y,x)$ with $\{ ab\}=\{tt, tx, ty, t\phi, xx, xy, x\phi,yy,y\phi, \phi\phi \}$.
After imposing the appropriate physical boundary conditions for unstable modes, the coupled system of the nine PDEs plus the algebraic equation for $\tilde{H}_{tt}$ yields an eigenvalue problem in the (complex) frequency $\widetilde{\omega}=\widetilde{\omega}_R+i\,\widetilde{\omega}_I$.

To discuss the boundary conditions\footnote{The reader interested on even more first principles discussions of boundary conditions for perturbations of a BH is invited to see e.g. \cite{Dias:2010maa,Dias:2015nua}.}, first note that  unstable modes have frequencies whose real part is smaller than the potential barrier height set by the spin-2 field mass, $\omega_R< \mu $. A Frobenius analysis at the essential singularity located at the asymptotic infinity $y=1$ finds that unstable modes must decay as
\begin{equation}\label{BC:y1}
\tilde{H}_{ab}\big|_{y\to 1}\sim e^{-\frac{\sqrt{\widetilde{\mu}^2-\widetilde{\omega}^2}}{1-y^2}}
(1-y^2)^{-n-\frac{\left(1+\widetilde{a}^2\right)\left( \widetilde{\mu}^2-2\widetilde{\omega}^2 \right)}{2\sqrt{\widetilde{\mu}^2-\widetilde{\omega}^2}}}\, , 
\end{equation}
where, as a boundary condition, we have already eliminated a solution that grows unbounded at infinity as $e^{\sqrt{\widetilde{\mu}^2-\widetilde{\omega}^2}/(1-y^2)}$ and one has
 $n=0$ for $\{ ab\}=\{ty, t\phi \}$, $n=1$ for $\{ ab\}=\{tx, yy,y\phi, \phi\phi \}$, $n=2$ for  $\{ ab\}=\{xy, x\phi \}$ and $n=3$ for  $\{ ab\}=\{xx \}$.

At the horizon, regularity of the $m=1$ perturbation in ingoing Eddington-Finkelstein coordinates requires that we impose the boundary condition,
\begin{equation}\label{BC:y0}
\tilde{H}_{ab}\big|_{y\to 0}\sim y^{-p-i\,\frac{\widetilde{\omega}-\widetilde{\Omega}_H}{2\pi \widetilde{T}_H}}\,,
\end{equation}
which effectively excludes outgoing modes, $\sim y^{i\,(\widetilde{\omega}-\widetilde{\Omega}_H)/(2\pi \widetilde{T}_H)}$, at the horizon. Here, $p=0$ for $\{ ab\}=\{ty,t\phi, yy,y\phi, \phi\phi \}$, $p=1$ for $\{ ab\}=\{tx,xy,x\phi\}$,  and $p=2$ for $\{ ab\}=\{ xx \}$.

Finally, we have to discuss the boundary conditions at north and south poles of the $S^2$, recalling that we are interested in $m=1$ modes. In these conditions, regularity requires that the perturbations behave as 
\begin{equation}\label{BC:x}
\tilde{H}_{ab}\big|_{x\to \pm 1}\sim (1-x^2)^{k}\,,
\end{equation}
which eliminates irregular modes that would diverge as $(1-x^2)^{-k}$, where $k=0$ for  $\{ ab\}=\{ty,xy \}$, $k=1$ for  $\{ ab\}=\{tx,t\phi, xx,x\phi, yy \}$, $k=2$ for $\tilde{H}_{y\phi}$ and  $k=3$ for $\tilde{H}_{\phi\phi}$.

The above boundary conditions \eqref{BC:y1}-\eqref{BC:x} are simply imposed if we introduce the ancillary functions $q_j$, ($j=1,\cdots,9$) defined has
%\lipsum[1]
\begin{widetext}
\begin{eqnarray}\label{def:qs}
&& \tilde{H}_{tx}(y,x)= 
y^{-1-i\,\frac{\widetilde{\omega}-\widetilde{\Omega}_H}{2\pi \widetilde{T}_H}}
(1-x^2)\,
e^{-\frac{\sqrt{\widetilde{\mu}^2-\widetilde{\omega}^2}}{1-y^2}}
(1-y^2)^{-1-\frac{\left(1+\widetilde{a}^2\right)\left( \widetilde{\mu}^2-2\widetilde{\omega}^2 \right)}{2\sqrt{\widetilde{\mu}^2-\widetilde{\omega}^2}}}
 \,q_1(y,x),\nonumber\\
&& \tilde{H}_{ty}(y,x)= 
y^{-i\,\frac{\widetilde{\omega}-\widetilde{\Omega}_H}{2\pi \widetilde{T}_H}}\,
e^{-\frac{\sqrt{\widetilde{\mu}^2-\widetilde{\omega}^2}}{1-y^2}}
(1-y^2)^{-\frac{\left(1+\widetilde{a}^2\right)\left( \widetilde{\mu}^2-2\widetilde{\omega}^2 \right)}{2\sqrt{\widetilde{\mu}^2-\widetilde{\omega}^2}}}
 \,q_2(y,x),\nonumber\\
 && \tilde{H}_{t\phi}(y,x)= 
y^{-i\,\frac{\widetilde{\omega}-\widetilde{\Omega}_H}{2\pi \widetilde{T}_H}}
(1-x^2)\,
e^{-\frac{\sqrt{\widetilde{\mu}^2-\widetilde{\omega}^2}}{1-y^2}}
(1-y^2)^{-\frac{\left(1+\widetilde{a}^2\right)\left( \widetilde{\mu}^2-2\widetilde{\omega}^2 \right)}{2\sqrt{\widetilde{\mu}^2-\widetilde{\omega}^2}}}
 \,q_3(y,x),\nonumber\\
 && \tilde{H}_{xx}(y,x)= 
 y^{-2-i\,\frac{\widetilde{\omega}-\widetilde{\Omega}_H}{2\pi \widetilde{T}_H}}
(1-x^2)\,
e^{-\frac{\sqrt{\widetilde{\mu}^2-\widetilde{\omega}^2}}{1-y^2}}
(1-y^2)^{-3-\frac{\left(1+\widetilde{a}^2\right)\left( \widetilde{\mu}^2-2\widetilde{\omega}^2 \right)}{2\sqrt{\widetilde{\mu}^2-\widetilde{\omega}^2}}}
 \,q_4(y,x),\nonumber\\
 && \tilde{H}_{xy}(y,x)= 
y^{-1-i\,\frac{\widetilde{\omega}-\widetilde{\Omega}_H}{2\pi \widetilde{T}_H}}\,
e^{-\frac{\sqrt{\widetilde{\mu}^2-\widetilde{\omega}^2}}{1-y^2}}
(1-y^2)^{-2-\frac{\left(1+\widetilde{a}^2\right)\left( \widetilde{\mu}^2-2\widetilde{\omega}^2 \right)}{2\sqrt{\widetilde{\mu}^2-\widetilde{\omega}^2}}}
 \,q_5(y,x),\nonumber\\
 && \tilde{H}_{x\phi}(y,x)= 
y^{-1-i\,\frac{\widetilde{\omega}-\widetilde{\Omega}_H}{2\pi \widetilde{T}_H}}
(1-x^2)\,
e^{-\frac{\sqrt{\widetilde{\mu}^2-\widetilde{\omega}^2}}{1-y^2}}
(1-y^2)^{-2-\frac{\left(1+\widetilde{a}^2\right)\left( \widetilde{\mu}^2-2\widetilde{\omega}^2 \right)}{2\sqrt{\widetilde{\mu}^2-\widetilde{\omega}^2}}}
 \,q_6(y,x),\nonumber\\
 && \tilde{H}_{yy}(y,x)= 
y^{-i\,\frac{\widetilde{\omega}-\widetilde{\Omega}_H}{2\pi \widetilde{T}_H}}
(1-x^2)\,
e^{-\frac{\sqrt{\widetilde{\mu}^2-\widetilde{\omega}^2}}{1-y^2}}
(1-y^2)^{-1-\frac{\left(1+\widetilde{a}^2\right)\left( \widetilde{\mu}^2-2\widetilde{\omega}^2 \right)}{2\sqrt{\widetilde{\mu}^2-\widetilde{\omega}^2}}}
 \,q_7(y,x),\nonumber\\
 && \tilde{H}_{y\phi}(y,x)= 
y^{-i\,\frac{\widetilde{\omega}-\widetilde{\Omega}_H}{2\pi \widetilde{T}_H}}
(1-x^2)^2\,
e^{-\frac{\sqrt{\widetilde{\mu}^2-\widetilde{\omega}^2}}{1-y^2}}
(1-y^2)^{-1-\frac{\left(1+\widetilde{a}^2\right)\left( \widetilde{\mu}^2-2\widetilde{\omega}^2 \right)}{2\sqrt{\widetilde{\mu}^2-\widetilde{\omega}^2}}}
 \,q_8(y,x),\nonumber\\
 && \tilde{H}_{\phi\phi}(y,x)= 
y^{-i\,\frac{\widetilde{\omega}-\widetilde{\Omega}_H}{2\pi \widetilde{T}_H}}
(1-x^2)^3\,
e^{-\frac{\sqrt{\widetilde{\mu}^2-\widetilde{\omega}^2}}{1-y^2}}
(1-y^2)^{-1-\frac{\left(1+\widetilde{a}^2\right)\left( \widetilde{\mu}^2-2\widetilde{\omega}^2 \right)}{2\sqrt{\widetilde{\mu}^2-\widetilde{\omega}^2}}}
 \,q_9(y,x)\,,
\end{eqnarray}
\end{widetext}
and search for the auxiliary functions $q_j$ that are smooth and that obey {\it derived} boundary conditions, i.e. conditions that follow simply from the equations of motion for $q_j$ evaluated at the boundaries, $y=0,1$ and $x=\pm 1$ (see  \cite{Dias:2015nua}). Further recall that  once the 9 functions $q_j$ are found, $\tilde{H}_{tt}$ is given by a simply function of $\{ \tilde{H}_{yy},\tilde{H}_{xx},\tilde{H}_{\phi\phi} \}$.

In more detail, at the horizon it follows from the equations of motion that $q_{1,2,3}$ satisfy inhomogeneous Dirichlet boundary conditions and $q_{4,5,6,7,8,9}$ obey homogeneous Neumann boundary conditions:
%\begin{widetext}
\begin{align}\label{BCqs:y0}
& q_1(0,x)=\frac{1-\alpha^2}{2\left(1+x^2(2-x^2)\alpha^2\right)}\,q_4(0,x)\,, \quad \nonumber \\
& q_2(0,x)=\frac{1-\alpha^2}{2\left(1+x^2(2-x^2)\alpha^2\right)}\,q_5(0,x)\,, \quad \nonumber \\
& q_3(0,x)=\frac{1-\alpha^2}{2\left(1+x^2(2-x^2)\alpha^2\right)}\,q_6(0,x)\,, \quad \nonumber \\
& \partial_y q_{4,5,6,7,8,9}(0,x)=0\,.
\end{align}
%\end{widetext}
At the asymptotic boundary, $q_{1,2,3,7}$ satisfy inhomogeneous Dirichlet boundary conditions
\begin{align}\label{BCqs:y1}
& 2 \widetilde{\omega}  q_1(1,x)+i \, \sqrt{\widetilde{\mu}^2-\widetilde{\omega}^2}\, q_4(1,x)=0\,, \nonumber \\
& 2 \widetilde{\omega}  q_2(1,x)+i \, \sqrt{\widetilde{\mu}^2-\widetilde{\omega}^2}\,  q_5(1,x)=0\,, \nonumber \\
& 2 \widetilde{\omega}  q_3(1,x)+i \, \sqrt{\widetilde{\mu}^2-\widetilde{\omega}^2}\,  q_6(1,x)=0\,, \nonumber \\
& \widetilde{\omega}^2 q_7(1,x)+ \frac{\widetilde{\mu}^2 q_4(1,x)+4 \widetilde{\omega}^2 q_9(1,x)}{2-x^2}=0\,,
\end{align}
and $q_{4,5,6,8,9}$ obey mixed (Robin) boundary conditions that are long and not enlightening to show (again they simply follow from the equations of motion at $y=1$).

Finally, at the poles $x=\pm 1$ of the sphere, the equations of motion give the derived boundary conditions:
\begin{align}\label{BCqs:x1}
& \partial_x q_j(\pm 1,y)=0\,,  \,, \quad\hbox{for $j=1,4,9$}  \nonumber \\
&\partial_x q_j(\pm 1,y)\mp q_j(\pm 1,y)=0\,, \quad\hbox{for $j=2,5,8$}  \nonumber \\
&q_3(\pm 1,y)\pm \frac{i \,q_2(\pm 1,y)}{2 \left(1+\alpha ^2 \left(1-y^2\right)^2\right)}=0\,, \nonumber \\
&q_6(\pm 1,y)\pm \frac{i \, q_5(\pm 1,y)}{2 \left(1+\alpha ^2 \left(1-y^2\right)^2\right)}=0\,,  \\
&q_7(\pm 1,y)\mp 4 i \left(1+\alpha ^2 \left(1-y^2\right)^2\right)^2 
\nonumber \\
& \hspace{2cm}\times \left(\frac{q_8(\pm 1,y)}{1+\alpha ^2 \left(1-y^2\right)^2}\mp i\,q_9(\pm 1,y)\right)=0\,. \nonumber
\end{align}

Summarizing, we have to solve a {\it nonlinear} eigenvalue problem $-$ e.g. $\widetilde{\omega}$ appears inside a square root in \eqref{BCqs:y1}  $-$ to find the eigenfrequencies $\widetilde{\omega}$ and associated eigenfunctions $q_j$ ($j=1,\cdots,9$) that obey the boundary conditions \eqref{BCqs:y0}$-$\eqref{BCqs:x1}. For that we use a Newton-Raphson root-finding algorithm introduced and discussed in detail in Sec.~III.C of the review \cite{Dias:2015nua}. The system of PDEs is then discretized using a pseudospectral collocation grid on Gauss-Chebyshev-Lobbato points in the compactified directions $y\in[0,1]$ and $x\in[-1,1]$ and the linear equations are solved using LU decomposition.
These numerical methods are very well tested. In particular they are the same that were used to compute the ultraspinning and bar-mode gravitational instabilities of rapidly spinning Myers-Perry BHs \cite{Dias:2009iu,Dias:2010maa,Dias:2010eu,Dias:2010gk,Dias:2011jg,Dias:2014eua}, the near-horizon scalar condensation and superradiant instabilities of BHs \cite{Dias:2010ma,Dias:2011tj}, the gravitational superradiant instability of Kerr-AdS BHs \cite{Dias:2013sdc,Cardoso:2013pza} and the electro-gravitational quasinormal modes of the Kerr-Newman BHs \cite{Dias:2015wqa}. This fact further gives confidence on the validity of the analysis and results of the present study. 

\section{Further numerical results}\label{app:2}

\begin{figure*}[t]
   \centering
    \includegraphics[width=0.96\textwidth]{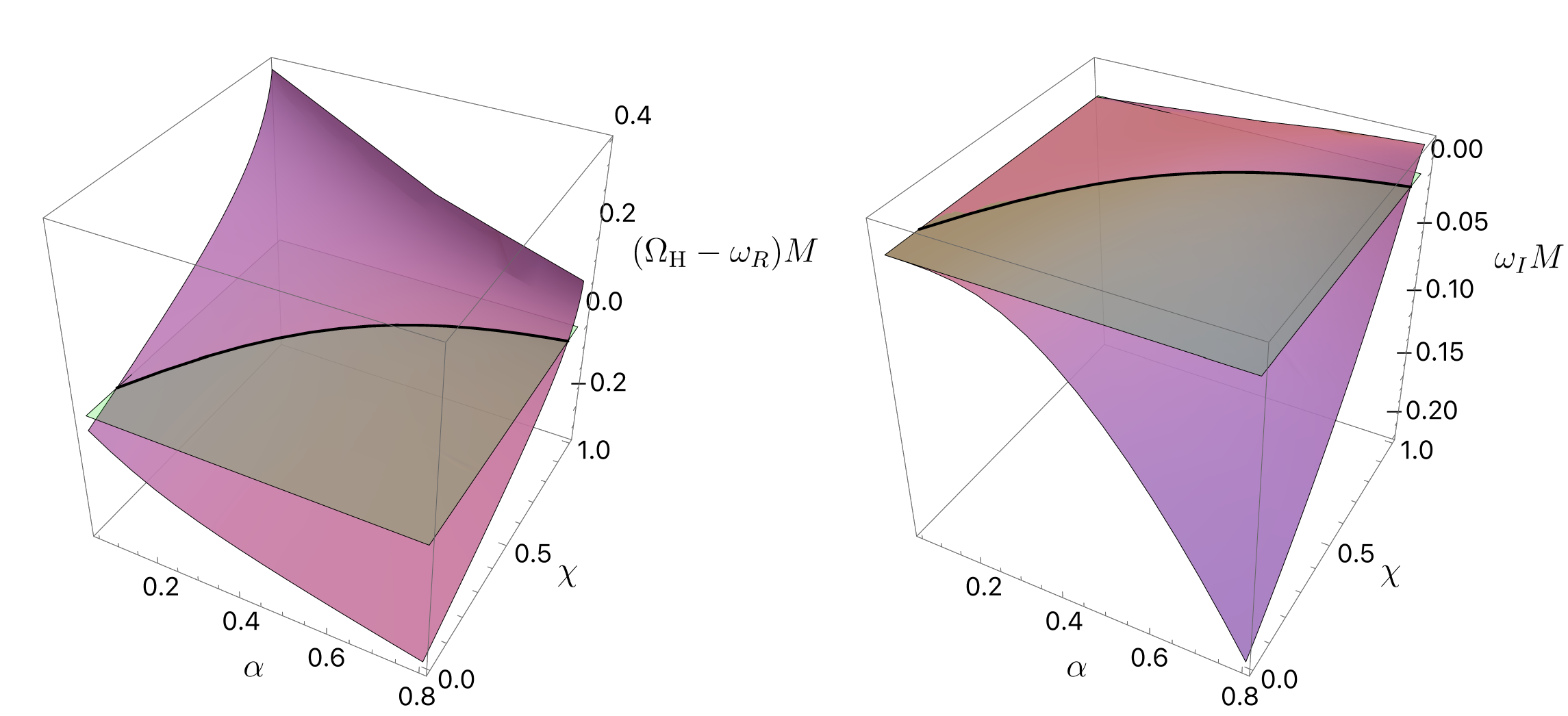}\\
    \includegraphics[width=0.96\textwidth]{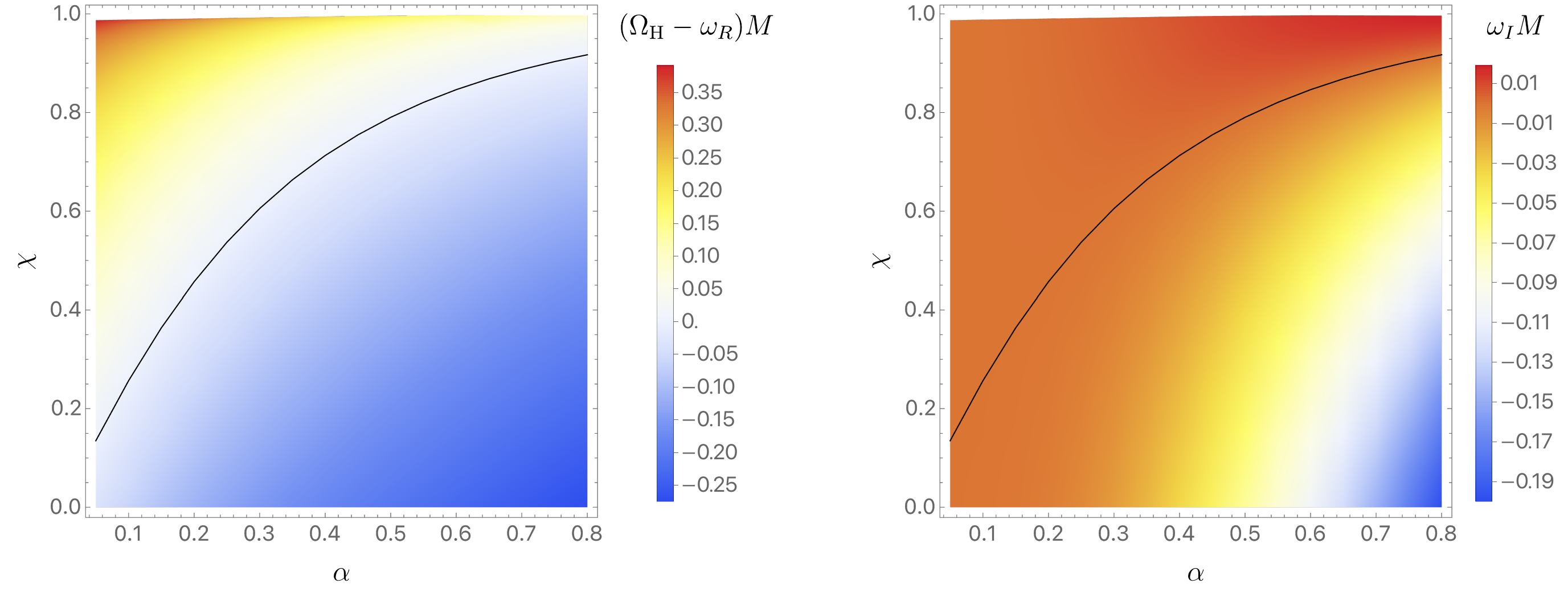}
    \caption{{\it Top panels:} Three-dimensional plot for the dominant unstable dipole mode ($m=1$) of massive spin-2 perturbations of a Kerr BH as functions of the coupling $\alpha\in[0.05,0.8]$ and BH spin $\chi\in[0,\approx 0.99]$. These plots were produced with 4018 data points. The solid black lines mark the onset of the instability, which also coincides with $\omega_R=\Omega_{\rm H}$, as expected for models exhibiting confining dynamics. To aid visualization we also added a gray plane marking $\Omega_{\rm H}-\omega_R=0$ and $\omega_I=0$.
    {\it Bottom panels:} Density plots of the same data.
    }
    \label{fig:3D}
\end{figure*}

In Fig.~\ref{fig:3D} we present a three-dimensional plot (top panels) and a contour plot (bottom panels) of the dominant unstable dipole  ($m=1$) mode discussed in the main text with all the data we collected. The right panels display the imaginary part of the frequency $\omega_I M$ as a function of the BH spin $\chi$ and coupling $\alpha$. The left panels display the difference between the angular velocity of the horizon and the real part of the frequency, $(\Omega_H-\omega_R)M$. The superradiant instability corresponds to $\Omega_H-\omega_R>0$ and $\omega_I>0$. 
In Fig.~1 of the main text we have taken the data of Fig.~\ref{fig:3D} corresponding to lines of constant $\alpha=0.1,\,0.2,\,0.3,\,0.4,\,0.5,\,0.6,\,0.7,\,0.8$.

We have validated our numerical method by:
\begin{itemize}
    \item Reproducing the analytical hydrogenic spectrum in the Newtonian ($\alpha\ll1$) limit~\cite{Brito:2020lup};
    \item Reproducing the numerical results for the special dipole mode in the nonspinning limit~\cite{Brito:2013wya};
    \item Verifying that all superradiant modes become marginally stable ($\omega_I=0$) precisely when the superradiant condition is saturated, $\omega=\Omega_H$, within the high numerical accuracy of our code. This instability onset is described by the black curve in all the plots of Fig.~\ref{fig:3D}. 
\end{itemize}

\bibliography{Refs.bib}

\end{document}